\documentclass[aps,prl,reprint,showpacs,floatfix]{revtex4-1}

\usepackage{graphicx}
\usepackage{amsmath,amssymb}
\usepackage{natbib}
\usepackage{siunitx}
\usepackage{color}

\begin{document}

\renewcommand{\theenumi}{(\roman{enumi})}%

\title{Mode-matching for Optical Antennas}
\author{Thorsten Feichtner}
\affiliation{Helmholtz-Zentrum Berlin f\"ur Materialien und Energie GmbH, Institut Nanoarchitekturen f\"ur die Energieumwandlung, Hahn-Meitner-Platz 1, 14109 Berlin, Germany}
\affiliation{Max Planck Institute for the Science of Light - Photonic Nanostructures, G\"unther-Scharowsky-Strasse 1, Bau 26, 91058 Erlangen, Germany}
\affiliation{Nano-Optics \& Biophotonics Group, Department of Experimental Physics 5,\\R\"ontgen Research Center for Complex Material Research (RCCM),\\Physics Institute, University of W\"urzburg, Am Hubland, D-97074 W\"urzburg, Germany}
\author{Silke Christiansen}
\affiliation{Helmholtz-Zentrum Berlin f\"ur Materialien und Energie GmbH, Institut Nanoarchitekturen f\"ur die Energieumwandlung, Hahn-Meitner-Platz 1, 14109 Berlin, Germany}
\affiliation{Max Planck Institute for the Science of Light - Photonic Nanostructures, G\"unther-Scharowsky-Strasse 1, Bau 26, 91058 Erlangen, Germany}
\affiliation{Freie Universit\"at Berlin, Arnimallee 14, 14195 Berlin}
\author{Bert Hecht}
\affiliation{Nano-Optics \& Biophotonics Group, Department of Experimental Physics 5,\\R\"ontgen Research Center for Complex Material Research (RCCM),\\Physics Institute, University of W\"urzburg, Am Hubland, D-97074 W\"urzburg, Germany}
\email{bert.hecht@physik.uni-wuerzburg.de}


\begin{abstract}
The emission rate of a point dipole can be strongly increased in presence of a well-designed optical antenna. Yet, optical antenna design is largely based on radio-frequency rules, ignoring e.g.~ohmic losses and non-negligible field penetration in metals at optical frequencies. Here we combine reciprocity and Poynting's theorem to derive a set of optical-frequency antenna design rules for benchmarking and optimizing the performance of optical antennas driven by single quantum emitters. Based on these findings a novel plasmonic cavity antenna design is presented exhibiting a considerably improved performance compared to a reference two-wire antenna. Our work will be useful for the design of high-performance optical antennas and nanoresonators for diverse applications ranging from quantum optics to antenna-enhanced single-emitter spectroscopy and sensing.
\end{abstract}

\maketitle

\section{Introduction}

Focusing optical antennas (FOAs) make use of plasmonic resonances to convert propagating electromagnetic waves at visible frequencies to near-fields localized in nanoscale volumes much smaller than the diffraction limit\cite{Novotny2011,Biagioni2012}. In such a hot spot the local density of states (LDOS) for point-like quantum emitters (QEs) may be increased by a factor of $10^3$ and possibly beyond \cite{Kinkhabwala2009,Biagioni2012,Akselrod2014}, which can be applied in novel light-based technologies, e.g.~quantum optics \cite{Torma2015} and communication \cite{Hoang2015}, sensing \cite{Stewart2008} as well as scanning near-field microscopy \cite{Novotny2006}. The design of FOAs, which typically consist of single or multiple particles of basic shapes \cite{Kalele2007,Hoang2015,Muhlschlegel2005,Kinkhabwala2009,Curto2010}, is largely inspired by rules derived from the radio frequency (rf) regime. The resulting antenna structures, however, can hardly be optimal for QE-FOA coupling, since there is no comparable task in rf-technology. In addition the radiation behavior of optical antennas differs from their rf-counterparts due to ohmic losses and fields penetrating the antenna material \cite{Dorfmuller2010}. Yet, it has been shown that the Purcell factor \cite{Sauvan2013,Kristensen2014} and likewise the antenna impedance \cite{Greffet2010} provide a measure for emitter-antenna coupling based on the antennas Green's function \cite{Krasnok2015}.   

Here we combine Poynting's theorem \cite{Hecht2012} with reciprocity \cite{Schwinger1998} to quantify QE-FOA coupling by means of a 3D overlap integral of the QE's electric field and the antenna's mode current pattern (cf.~mode matching \cite{Snyder1983,Then2014}). Introducing a further mode-matching condition for FOA to far-field coupling allows us to identify two independent FOA mode current patterns, which both maximize antenna radiation. This enables us to understand the high performance of FOAs obtained from evolutionary optimization \cite{Feichtner2015} as well as of other unusual FOA geometries, like the indented nano-cone \cite{Garcia-Etxarri2012} or the double hole resonator \cite{Regmi2015}. Finally, based on our new design rules, an improved plasmonic cavity antenna geometry is devised and numerically investigated. The flexibility of the presented framework opens diverse applications ranging from improved emitter-cavity coupling in quantum optics to enhanced single-emitter sensing schemes. It also provides new insights for the understanding and optimization of complex-shaped metal nano-objects as they appear in surface-enhanced Raman scattering substrates \cite{Schlucker2014}.

\section{Theory}

We consider a point dipole with dipole moment $\mathbf{p}$ situated at $\mathbf{r}_p$, emitting photons at wavenumber $k$ with unity quantum efficiency. The emitted power $P$ of the dipole in an arbitrary environment depends on the self-interaction due to scattered fields $\mathbf{E}_\text{sc}$. The enhancement of QE emission rate $\gamma / \gamma_0$ as well as of the dipole emission power $P / P_0$ in an inhomogeneous environment can be calculated based on Poynting's theorem \cite{Hecht2012}:
\begin{equation}
\frac{\gamma}{\gamma_0} = \frac{P}{P_0} = 1 + \frac{6 \pi \varepsilon_0}{\left| \mathbf{p} \right|^2} \frac{1}{k^3} \text{Im}\, \left \{ \mathbf{E}_\text{sc}(\mathbf{r}_p) \cdot \mathbf{p}^* \right \}\, .
\label{eq:dipTotalPower}
\end{equation}
Here $\gamma_0$ and $P_0$ are the vacuum values of the QE emission rate and dipole emission power, respectively. The emission power enhancement depends on the backscattered field components at the dipole position parallel to the dipole moment. Equation \eqref{eq:dipTotalPower} also takes into account the phase between dipolar moment and scattered field $\Delta \phi = \phi_\text{sc} - \phi_\text{p}$ \cite{Drexhage1974}
\begin{equation}
\text{Im}\left\{ \mathbf{E}_\text{sc}(\mathbf{r}_p) \cdot \mathbf{p}^* \right\} =  \left | \mathbf{E}_\text{sc}(\mathbf{r}_p)\right | \cdot  \left | \mathbf{p} \right | \cdot \text{Im}\left\{ e^{i\, \Delta \phi} \right\}\, .
\label{eq:phaseShift}
\end{equation}

\begin{figure}[tp]
	\centering
		\includegraphics{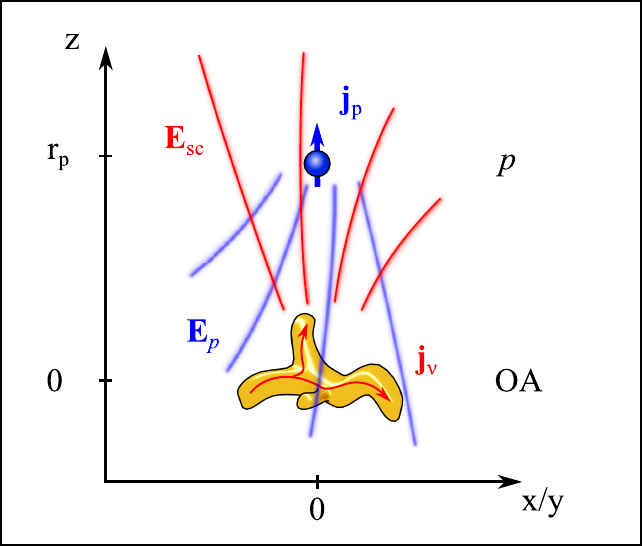}
	\caption{(color online) General setup of a dipole $p$ situated at $\mathbf{r}_p = (0,0,r_p)$ with an oscillating current $\mathbf{j}_p$ being the source of electromagnetic fields $\mathbf{E}_p$. In its environment a metallic nanoparticle is situated with a single excitable mode $\nu$ leading to scattered fields $\mathbf{E}_{p/\nu}$ originating from source current densities $\mathbf{j}_{p/\nu}$.}
	\label{fig:gen_setup}
\end{figure}
We define the scattering environment to be a general FOA with its center of mass in the coordinate origin as sketched in Fig.~\ref{fig:gen_setup}. It exhibits a set of plasmonic eigenmodes at the emission wavelength of the QE which is positioned on the $z$-axis at $\mathbf{r}_p = (0,0,r_p)$ with $k\cdot r_p \ll 1$ and its dipole moment oriented along $z$. Its current density then can be written as $j_p=-i\omega\mathbf{p}\,\delta\left( \mathbf{r} - \mathbf{r}_p \right)$. In the following we assume without loss of generality only a single relevant FOA mode $\nu$. The scattered field in equation \eqref{eq:dipTotalPower} can then be expressed as:
\begin{equation}
\mathbf{E}_\text{sc}(\mathbf{r}_p) = i \omega \mu_0 \int_{V_\nu} \bar{\mathbf{G}}_0(\mathbf{r}_p,\mathbf{r}') \mathbf{j}_\nu (\mathbf{r}') \, \text{d}^3 r'\; ,
\label{eq:def_Esc}
\end{equation}
where $\bar{\mathbf{G}}_0(\mathbf{r}_p,\mathbf{r}')$ is the Green's tensor and $\mathbf{j}_\nu$ is the modes current density being the source of the scattered field. For a FOA consisting of a local, dispersive and lossy material, which is described by the dielectric function $\varepsilon (\omega)$, the reciprocity theorem implies the symmetry $\bar{\mathbf{G}}(\mathbf{r}_p,\mathbf{r}') = \bar{\mathbf{G}}(\mathbf{r}',\mathbf{r}_p)$ of the Green's tensor \cite{Schwinger1998}. Inserted into \eqref{eq:def_Esc} the scattered fields now depend on the Green's function of the emitting dipole at $\mathbf{r}_p$ evaluated inside the volume of the FOA:
\begin{equation}
\mathbf{E}_\text{sc}(\mathbf{r}_p) = i \omega \mu_0 \int_{V_\nu} \bar{\mathbf{G}}_0(\mathbf{r}',\mathbf{r}_p) \mathbf{j}_\nu (\mathbf{r}') \, \text{d}^3 r'\, .
\label{eq:Scat_reci}
\end{equation}
Exciting the FOA at resonance in the quasistatic limit leads to $\Delta \phi = \pi / 2$ and thus to $\text{Im}\left\{ e^{i\, \Delta \phi} \right\} = 1$. Together with \eqref{eq:Scat_reci} in \eqref{eq:dipTotalPower} this leads to:
\begin{multline}
\frac{\gamma}{\gamma_0} = \frac{P}{P_0} = 1 + \frac{6 \pi \varepsilon_0}{\left| \mathbf{p} \right|^2} \frac{1}{k^3} \cdot \\ \cdot \omega \mu_0 \int_{V_\nu} \left | \bar{\mathbf{G}}_0(\mathbf{r}',\mathbf{r}_p) \, \mathbf{p} \, \mathbf{j}_\nu (\mathbf{r}')\right | \, \text{d}^3 r'(\mathbf{r}_p)\; .
\label{eq:dipTotalPower_GInt}
\end{multline}
Using $\mathbf{E}_p(\mathbf{r}) = \omega^2 \mu_0 \bar{\mathbf{G}}(\mathbf{r},\mathbf{r}_p)\, \mathbf{p}$ we obtain the important result:
\begin{equation}
\frac{P}{P_0} = 1 + \frac{6 \pi c \varepsilon_0}{k^4} \int_{V_\nu} \left | \mathbf{E}_p (\mathbf{r}') \cdot \mathbf{j}_\nu (\mathbf{r}')\right | \, \text{d}^3 r'\; ,
\label{eq:dipTotalPower_reci}
\end{equation}
with $c$ the speed of light in vacuum. This equation describes the fact that the total power extracted from a point dipole into the $\nu$-th antenna mode (i.e.~the Purcell factor) is described by the overlap integral of the mode's current density pattern with the point dipole fields inside the volume of the FOA, thereby defining a mode-matching condition. To test the validity of this equation, the analytical case of a dipole in front of a sphere has been evaluated (Appendix I) and a numerical test on split-ring antennas has been performed (Appendix II).

\section{Discussion}

Equation \eqref{eq:dipTotalPower_reci} is reminiscent of mode matching formalisms used to determine coupling efficiencies between waveguide modes \cite{Snyder1983,Then2014}. Here, however, the three-dimensional volume of the FOA has to be considered. From \eqref{eq:dipTotalPower_reci} three intuitive rules can be deduced: (i) Align the dipole field and the antenna's mode current pattern everywhere inside the antenna volume. (ii) Maximize the mode current at each point inside the plasmonic antenna. This is mainly a material issue as $\mathbf{j} = \sigma \cdot \mathbf{E}_\nu$ depends on the conductivity $\sigma$. (iii) Maximize the volume of the overlap integral. This suggest the use of as much as possible metal in the vicinity of the dipole. Rules (i) and (iii) suggest that the established two-particle geometries may not result in the best possible FOAs. Instead, a FOA should enclose the QE as much as possible, resembling a kind of plasmonic cavity antenna. Antennas that to some extend fulfill these design rules in two dimensions have already been realized and are known as double-hole resonators\cite{Regmi2015}.

\begin{figure*}[tp]
	\centering
		\includegraphics[width=\textwidth]{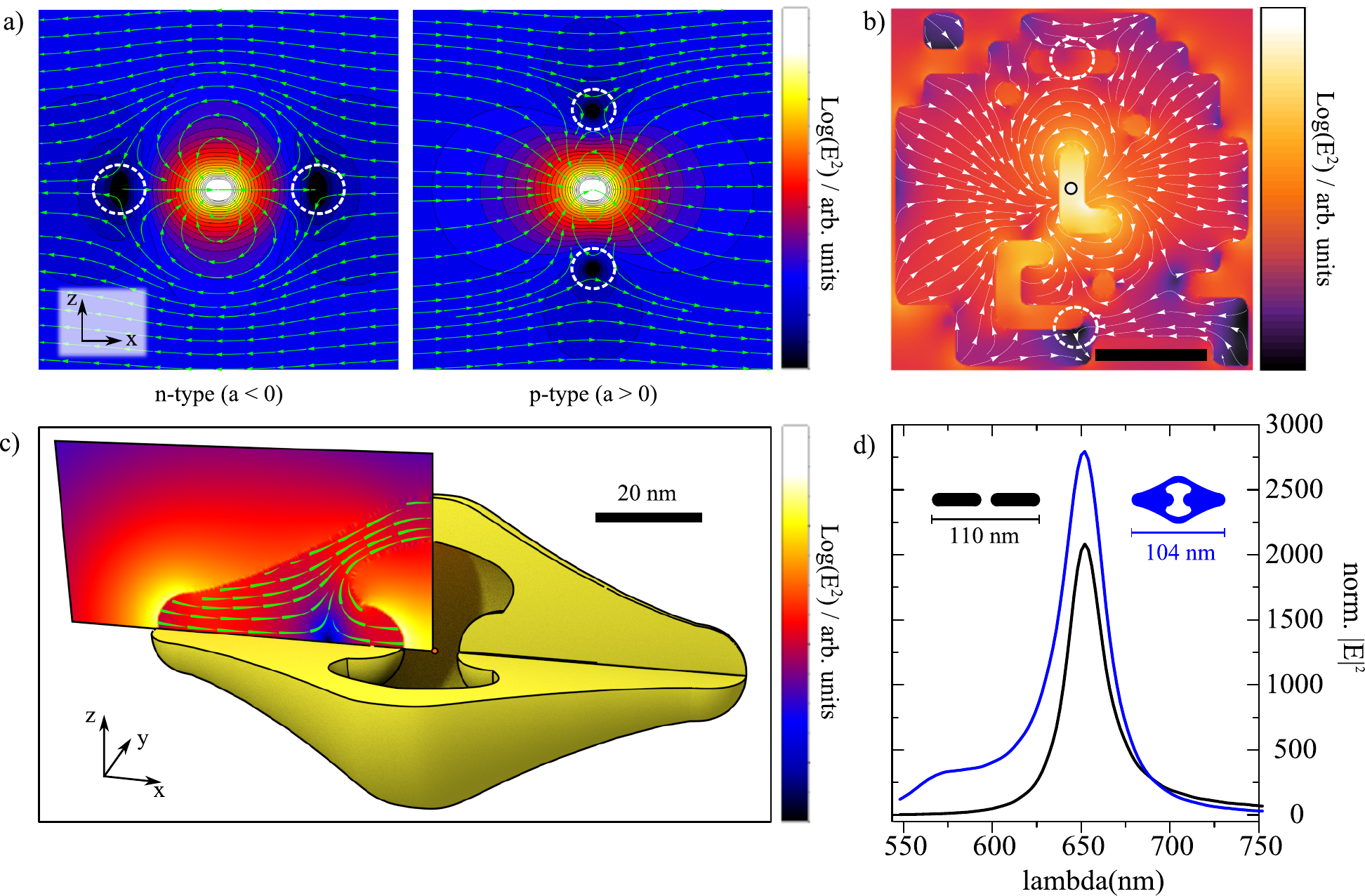}
	\caption{(color online) Plasmonic antenna modes with double mode-matching. (a) $x$-$y$-cross section of the linear combination of quasistatic dipolar field and constant field pointing in $x$-direction as described by equation \eqref{eq:mode_field_blueprint} for $a<0$ (left) and $a>0$ (right). The white dashed circles mark point of vanishing fields for better orientation in panel (b), showing the near field intensity enhancement (color scale) and the current direction (white arrows) of a planar antenna geometry optimized an evolutionary algorithm for maximum fields in the center (marked by small circle; scale-bar 100 nm)\cite{Feichtner2015}. (c) Antenna design carrying a resonant mode resembling panel (a) left. The originally rotational symmetric geometry is shown with a 90° cutaway for improved visualization. Additionally the near-field intensity (color) as well as the current direction (green arrows) are overlay for a quarter cross section. The small orange dot marks the center, where a QE is to be placed. (d) Near-field intensity enhancement spectra at the green point in panel (c) (blue) as well as at the center of a two wire dipole reference antenna, with identical end cap radius (black). The small insets show a $x$-$y$-plane cross section of both geometries.}
	\label{fig:int_design}
\end{figure*}

So far we have only considered the transfer of energy from the dipole to the antenna mode. However, a FOA has two tasks, which have to been fulfilled by the antenna mode field $\mathbf{E}_\nu = \mathbf{j}_\nu \cdot \sigma^{-1}$ (see \cite{Olmon2012}): In addition to providing a maximal LDOS at the emitters position it also has to couple efficiently to propagating far-fields. This can be modeled by applying mode-matching according to \eqref{eq:dipTotalPower_reci} again considering a dipole in the far-field oriented parallel to the first dipole. Since the electric fields of the second dipole at the FOA are plane waves, in order to optimize far-field coupling, a homogeneous field $\mathbf{E}_c$ has to be added to the dipolar near fields originating from the QE close to the antenna. For a QE oriented along the $x$-axis the optimal mode current pattern to fulfill both mode-matching criteria is therefore a linear combination of quasi-static dipolar contribution:
\begin{equation}
\mathbf{E}_d = \frac{1}{4\pi \varepsilon_0} \frac{3\mathbf{n}\left(\mathbf{n}\mathbf{p}\right) - \mathbf{p}}{r^3}\; ,
\label{eq:qsDipole}
\end{equation}
with the homogeneous field $\mathbf{E}_c = a \cdot \hat{\mathbf{x}}$:
\begin{equation}
\mathbf{E}_\nu = \mathbf{E}_d + \mathbf{E}_c \quad .
\label{eq:mode_field_blueprint}
\end{equation}
By assuming a pure dipolar field we omit the refraction of the fields at the antenna surface. This is a good approximation if the antenna material strongly deviates from an ideal metal (for discussion see Appendix III). The scalar factor $a$ can be positive or negative, leading to two fundamentally different optimal focusing antenna mode current patterns as illustrated in Fig.~\ref{fig:int_design}(a), which we denote as 'n-type' (left) and 'p-type' (right) mode. Close to the antenna hotspot in which the QE is positioned the dipolar near-field $\mathbf{E}_d$ dominates. Away from the QE, $\mathbf{E}_d$ falls off as $1/r^3$ and the homogeneous field starts to dominate. For the n-type mode (p-type mode) isolated points on the $x$-axis (a continuous circle in the $y$-$z$-plane) with zero field strength appear (marked with white dashed circles).

Figure~\ref{fig:int_design}(b) shows a mode current pattern of a FOA that resulted from an evolutionary algorithm \cite{Feichtner2015} with the optimization goal to maximize the near-field in in the center of the structure using a planar 30 nm thick patterned gold layer at $\lambda=830$ nm illuminated by a horizontally polarized focused Gaussian beam. The current pattern is identified as a p-type mode (areas with vanishing fields marked with white dashed line): The antenna center is surrounded by gold, realizing the 2D-equivalent of a plasmonic cavity antenna. The currents switch direction on the $y$-axis to match the needs for optimal far-field coupling. Since the antenna is sufficiently small and centered in a Gaussian focus, plane wave excitation can be assumed. While a p-type mode current pattern can be obtained in this 2D arrangement it cannot be realized with benefits in 3D as discussed in Appendix IV.

Figure~\ref{fig:int_design}(c) introduces a 3D-plasmonic cavity antenna supporting the n-type mode current pattern, which to our knowledge has not yet been realized in an optical antenna \footnote{The mode and the spectra were obtained numerically using Lumerical FDTD Solutions, using focussed broadband ($\lambda = 550 - 750$ nm) Gaussian illumination ($NA=1$, $3.9$ fs pulse length). The gold material was implemented according to\cite{Etchegoin2006}}. We choose a geometry with rotational symmetry based on a reference antenna (two-wire dipole antenna, 10 nm gap, 15 nm wire diameter, spherical end caps, overall length $l=110$ nm made from gold, see small black inset in Fig.~\ref{fig:int_design}(d)). The reference antenna has a resonance at $\lambda = 650$ nm (black graph, Fig.~\ref{fig:int_design}(d)). To realize the plasmonic cavity antenna, interconnects were attached between the antenna arms to allow additional current paths, enclosing the QE with gold. The length of the plasmonic cavity antenna was tuned to also be resonant at $\lambda=650$ nm resulting in a slightly reduced length of $l=104$ nm,  5.5\% shorter than the reference antenna. The plasmonic cavity antenna is a single particle with a mode current pattern flowing uni-directionally from end to end resembling a $\lambda/2$-resonance (see inset of Fig.~\ref{fig:int_design}(c)). The mode current pattern also exhibits areas of vanishing fields along the $x$-axis, as expected for an n-type mode (left panel of Fig. \ref{fig:int_design}(a)). In contrary, the reference antenna exhibits a $\lambda$-resonance \cite{Biagioni2012} similar to voltage-fed radio-frequency antennas \cite{Lee1984}.

Figure \ref{fig:int_design}(d) shows the near-field intensity enhancement spectra at the antenna center for both plasmonic cavity antenna and reference antenna when illuminated by a Gaussian focus with a numerical aperture of \textit{NA}$\,=1$. Both antenna resonances peak at $\lambda=650$ nm with a near-field intensity enhancement of $2.79\cdot 10^3$ ($2.08\cdot 10^3$) for the plasmonic cavity antenna (reference antenna) and a Q-factor $Q=\lambda/\Delta\lambda$ of $Q =22.0$ ($Q =27.1$). The plasmonic cavity antenna spectrum shows a second shallow peak at 570 nm, due to a mode similar to that of the reference antenna with a current density in $x$-direction that does not change its direction. An analysis of the antenna cross sections under plane wave illumination was performed to identify the additional loss channels of the plasmonic cavity antenna that lead to the 19\% decreased Q-factor. The plasmonic cavity antenna exhibits an absorption cross section of $4.17\cdot 10^4$ nm$^2$ and a scattering cross section of $1.77\cdot 10^4$ nm$^2$, yielding a scattering efficiency of $\eta = 0.298$. The reference antenna exhibits an absorption cross section of $2.34\cdot 10^4$ nm$^2$ and a scattering cross section of $0.362\cdot 10^4$ nm$^2$ and, thus, a 55\% lower scattering efficiency of $\eta = 0.134$. Thus, the lowered Q-factor for the plasmonic cavity antenna is due to increased radiation losses. According to \cite{Tretyakov2014}, particles small compared to the impinging wavelength couple best to the far-field, when both scattering and absorption cross section are equal. Also, the plasmonic cavity antenna mode currents from end to end resemble that of a $\lambda/2$-antenna, which in rf-technology is known to radiate most efficiently\cite{Lee1984}.

Although realizing an n-type mode current pattern, the plasmonic cavity antenna design in Fig.~\ref{fig:int_design}(c) likely does not represent the ultimate limit that can be achieved in terms of near-field intensity enhancement since it has been designed to be as similar as possible to the reference antenna. To find better designs and the optimal magnitude of $a$ in equation \eqref{eq:mode_field_blueprint} will be a topic of future research.

\section{Conclusion}

Reciprocity and Poyntings theorem can be combined to obtain a three-dimensional mode-matching framework for describing optimal coupling between a quantum emitter and a plasmonic optical antenna. Based on this framework we identified two fundamental mode current patterns for optimal focusing optical antennas. Making use of these new design rules for optical antennas the concept of the plasmonic cavity antenna has been developed which outperforms a two-wire reference antenna.

The developed framework will help to unravel the full potential of focusing optical antennas and will help to optimize e.g.~TERS-tips and SERS substrates, making use of novel complex and surprising geometries \cite{Garcia-Etxarri2012,Regmi2015}. Both near-field and far-field coupling tasks of an optical antenna can be optimized rather independently yielding a large flexibility to optimize antenna performance for a variety of different tasks. An extension to multi-particle multi-mode systems with retardation, where several double mode-matching conditions define the overall far-field to near-field conversion efficiency will help to better understand complex and large scale SERS-active substrates\cite{Schlucker2014}. On the other hand, by deliberately avoiding far-field coupling and concentrating on the quantum emitter to antenna coupling only \cite{Akselrod2014}, nano cavities for strong light matter coupling can be devised.

\section{Acknowledgements}
TF acknowledges Katja Hoeflich for stimulating discussions. The work was funded by the DFG (HE5648/1-1 and CRC951).

\bibliography{reci_ant_paper}

\begin{thebibliography}{40}%
\makeatletter
\providecommand \@ifxundefined [1]{%
 \@ifx{#1\undefined}
}%
\providecommand \@ifnum [1]{%
 \ifnum #1\expandafter \@firstoftwo
 \else \expandafter \@secondoftwo
 \fi
}%
\providecommand \@ifx [1]{%
 \ifx #1\expandafter \@firstoftwo
 \else \expandafter \@secondoftwo
 \fi
}%
\providecommand \natexlab [1]{#1}%
\providecommand \enquote  [1]{``#1''}%
\providecommand \bibnamefont  [1]{#1}%
\providecommand \bibfnamefont [1]{#1}%
\providecommand \citenamefont [1]{#1}%
\providecommand \href@noop [0]{\@secondoftwo}%
\providecommand \href [0]{\begingroup \@sanitize@url \@href}%
\providecommand \@href[1]{\@@startlink{#1}\@@href}%
\providecommand \@@href[1]{\endgroup#1\@@endlink}%
\providecommand \@sanitize@url [0]{\catcode `\\12\catcode `\$12\catcode
  `\&12\catcode `\#12\catcode `\^12\catcode `\_12\catcode `\%12\relax}%
\providecommand \@@startlink[1]{}%
\providecommand \@@endlink[0]{}%
\providecommand \url  [0]{\begingroup\@sanitize@url \@url }%
\providecommand \@url [1]{\endgroup\@href {#1}{\urlprefix }}%
\providecommand \urlprefix  [0]{URL }%
\providecommand \Eprint [0]{\href }%
\providecommand \doibase [0]{http://dx.doi.org/}%
\providecommand \selectlanguage [0]{\@gobble}%
\providecommand \bibinfo  [0]{\@secondoftwo}%
\providecommand \bibfield  [0]{\@secondoftwo}%
\providecommand \translation [1]{[#1]}%
\providecommand \BibitemOpen [0]{}%
\providecommand \bibitemStop [0]{}%
\providecommand \bibitemNoStop [0]{.\EOS\space}%
\providecommand \EOS [0]{\spacefactor3000\relax}%
\providecommand \BibitemShut  [1]{\csname bibitem#1\endcsname}%
\let\auto@bib@innerbib\@empty
\bibitem [{\citenamefont {Novotny}\ and\ \citenamefont {van
  Hulst}(2011)}]{Novotny2011}%
  \BibitemOpen
  \bibfield  {author} {\bibinfo {author} {\bibfnamefont {L.}~\bibnamefont
  {Novotny}}\ and\ \bibinfo {author} {\bibfnamefont {N.}~\bibnamefont {van
  Hulst}},\ }\href {\doibase 10.1038/nphoton.2010.237} {\bibfield  {journal}
  {\bibinfo  {journal} {Nature Photonics}\ }\textbf {\bibinfo {volume} {5}},\
  \bibinfo {pages} {83} (\bibinfo {year} {2011})}\BibitemShut {NoStop}%
\bibitem [{\citenamefont {Biagioni}\ \emph {et~al.}(2012)\citenamefont
  {Biagioni}, \citenamefont {Huang},\ and\ \citenamefont
  {Hecht}}]{Biagioni2012}%
  \BibitemOpen
  \bibfield  {author} {\bibinfo {author} {\bibfnamefont {P.}~\bibnamefont
  {Biagioni}}, \bibinfo {author} {\bibfnamefont {J.-S.}\ \bibnamefont {Huang}},
  \ and\ \bibinfo {author} {\bibfnamefont {B.}~\bibnamefont {Hecht}},\
  }\href@noop {} {\bibfield  {journal} {\bibinfo  {journal} {Rep. Prog. Phys.}\
  }\textbf {\bibinfo {volume} {75}},\ \bibinfo {pages} {024402} (\bibinfo
  {year} {2012})}\BibitemShut {NoStop}%
\bibitem [{\citenamefont {Kinkhabwala}\ \emph {et~al.}(2009)\citenamefont
  {Kinkhabwala}, \citenamefont {Yu}, \citenamefont {Fan}, \citenamefont
  {Avlasevich}, \citenamefont {M\"{u}llen}, \citenamefont {Moerner},\ and\
  \citenamefont {Muellen}}]{Kinkhabwala2009}%
  \BibitemOpen
  \bibfield  {author} {\bibinfo {author} {\bibfnamefont {A.}~\bibnamefont
  {Kinkhabwala}}, \bibinfo {author} {\bibfnamefont {Z.}~\bibnamefont {Yu}},
  \bibinfo {author} {\bibfnamefont {S.}~\bibnamefont {Fan}}, \bibinfo {author}
  {\bibfnamefont {Y.}~\bibnamefont {Avlasevich}}, \bibinfo {author}
  {\bibfnamefont {K.}~\bibnamefont {M\"{u}llen}}, \bibinfo {author}
  {\bibfnamefont {W.~E.}\ \bibnamefont {Moerner}}, \ and\ \bibinfo {author}
  {\bibfnamefont {K.}~\bibnamefont {Muellen}},\ }\href {\doibase
  10.1038/nphoton.2009.187} {\bibfield  {journal} {\bibinfo  {journal} {Nature
  Photonics}\ }\textbf {\bibinfo {volume} {3}},\ \bibinfo {pages} {654}
  (\bibinfo {year} {2009})}\BibitemShut {NoStop}%
\bibitem [{\citenamefont {Akselrod}\ \emph {et~al.}(2014)\citenamefont
  {Akselrod}, \citenamefont {Argyropoulos}, \citenamefont {Hoang},
  \citenamefont {Cirac\`{\i}}, \citenamefont {Fang}, \citenamefont {Huang},
  \citenamefont {Smith},\ and\ \citenamefont {Mikkelsen}}]{Akselrod2014}%
  \BibitemOpen
  \bibfield  {author} {\bibinfo {author} {\bibfnamefont {G.~M.}\ \bibnamefont
  {Akselrod}}, \bibinfo {author} {\bibfnamefont {C.}~\bibnamefont
  {Argyropoulos}}, \bibinfo {author} {\bibfnamefont {T.~B.}\ \bibnamefont
  {Hoang}}, \bibinfo {author} {\bibfnamefont {C.}~\bibnamefont {Cirac\`{\i}}},
  \bibinfo {author} {\bibfnamefont {C.}~\bibnamefont {Fang}}, \bibinfo {author}
  {\bibfnamefont {J.}~\bibnamefont {Huang}}, \bibinfo {author} {\bibfnamefont
  {D.~R.}\ \bibnamefont {Smith}}, \ and\ \bibinfo {author} {\bibfnamefont
  {M.~H.}\ \bibnamefont {Mikkelsen}},\ }\href {\doibase
  10.1038/nphoton.2014.228} {\bibfield  {journal} {\bibinfo  {journal} {Nature
  Photonics}\ }\textbf {\bibinfo {volume} {8}},\ \bibinfo {pages} {835}
  (\bibinfo {year} {2014})}\BibitemShut {NoStop}%
\bibitem [{\citenamefont {T\"{o}rm\"{a}}\ and\ \citenamefont
  {Barnes}(2015)}]{Torma2015}%
  \BibitemOpen
  \bibfield  {author} {\bibinfo {author} {\bibfnamefont {P.}~\bibnamefont
  {T\"{o}rm\"{a}}}\ and\ \bibinfo {author} {\bibfnamefont {W.~L.}\ \bibnamefont
  {Barnes}},\ }\href {\doibase 10.1088/0034-4885/78/1/013901} {\bibfield
  {journal} {\bibinfo  {journal} {Reports on Progress in Physics}\ }\textbf
  {\bibinfo {volume} {78}},\ \bibinfo {pages} {013901} (\bibinfo {year}
  {2015})}\BibitemShut {NoStop}%
\bibitem [{\citenamefont {Hoang}\ \emph {et~al.}(2015)\citenamefont {Hoang},
  \citenamefont {Akselrod}, \citenamefont {Argyropoulos}, \citenamefont
  {Huang}, \citenamefont {Smith},\ and\ \citenamefont {Mikkelsen}}]{Hoang2015}%
  \BibitemOpen
  \bibfield  {author} {\bibinfo {author} {\bibfnamefont {T.~B.}\ \bibnamefont
  {Hoang}}, \bibinfo {author} {\bibfnamefont {G.~M.}\ \bibnamefont {Akselrod}},
  \bibinfo {author} {\bibfnamefont {C.}~\bibnamefont {Argyropoulos}}, \bibinfo
  {author} {\bibfnamefont {J.}~\bibnamefont {Huang}}, \bibinfo {author}
  {\bibfnamefont {D.~R.}\ \bibnamefont {Smith}}, \ and\ \bibinfo {author}
  {\bibfnamefont {M.~H.}\ \bibnamefont {Mikkelsen}},\ }\href {\doibase
  10.1038/ncomms8788} {\bibfield  {journal} {\bibinfo  {journal} {Nature
  communications}\ }\textbf {\bibinfo {volume} {6}},\ \bibinfo {pages} {7788}
  (\bibinfo {year} {2015})}\BibitemShut {NoStop}%
\bibitem [{\citenamefont {Stewart}\ \emph {et~al.}(2008)\citenamefont
  {Stewart}, \citenamefont {Anderton}, \citenamefont {Thompson}, \citenamefont
  {Maria}, \citenamefont {Gray}, \citenamefont {Rogers},\ and\ \citenamefont
  {Nuzzo}}]{Stewart2008}%
  \BibitemOpen
  \bibfield  {author} {\bibinfo {author} {\bibfnamefont {M.~E.}\ \bibnamefont
  {Stewart}}, \bibinfo {author} {\bibfnamefont {C.~R.}\ \bibnamefont
  {Anderton}}, \bibinfo {author} {\bibfnamefont {L.~B.}\ \bibnamefont
  {Thompson}}, \bibinfo {author} {\bibfnamefont {J.}~\bibnamefont {Maria}},
  \bibinfo {author} {\bibfnamefont {S.~K.}\ \bibnamefont {Gray}}, \bibinfo
  {author} {\bibfnamefont {J.~A.}\ \bibnamefont {Rogers}}, \ and\ \bibinfo
  {author} {\bibfnamefont {R.~G.}\ \bibnamefont {Nuzzo}},\ }\href@noop {}
  {\bibfield  {journal} {\bibinfo  {journal} {Chemical Reviews}\ }\textbf
  {\bibinfo {volume} {108}},\ \bibinfo {pages} {494} (\bibinfo {year}
  {2008})}\BibitemShut {NoStop}%
\bibitem [{\citenamefont {Novotny}\ and\ \citenamefont
  {Stranick}(2006)}]{Novotny2006}%
  \BibitemOpen
  \bibfield  {author} {\bibinfo {author} {\bibfnamefont {L.}~\bibnamefont
  {Novotny}}\ and\ \bibinfo {author} {\bibfnamefont {S.~J.}\ \bibnamefont
  {Stranick}},\ }\href {\doibase 10.1146/annurev.physchem.56.092503.141236}
  {\bibfield  {journal} {\bibinfo  {journal} {Annual review of physical
  chemistry}\ }\textbf {\bibinfo {volume} {57}},\ \bibinfo {pages} {303}
  (\bibinfo {year} {2006})}\BibitemShut {NoStop}%
\bibitem [{\citenamefont {Kalele}\ \emph {et~al.}(2007)\citenamefont {Kalele},
  \citenamefont {Tiwari}, \citenamefont {Gosavi},\ and\ \citenamefont
  {Kulkarni}}]{Kalele2007}%
  \BibitemOpen
  \bibfield  {author} {\bibinfo {author} {\bibfnamefont {S.~A.}\ \bibnamefont
  {Kalele}}, \bibinfo {author} {\bibfnamefont {N.~R.}\ \bibnamefont {Tiwari}},
  \bibinfo {author} {\bibfnamefont {S.~W.}\ \bibnamefont {Gosavi}}, \ and\
  \bibinfo {author} {\bibfnamefont {S.~K.}\ \bibnamefont {Kulkarni}},\
  }\href@noop {} {\bibfield  {journal} {\bibinfo  {journal} {Journal of
  Nanophotonics}\ }\textbf {\bibinfo {volume} {1}},\ \bibinfo {pages} {1}
  (\bibinfo {year} {2007})}\BibitemShut {NoStop}%
\bibitem [{\citenamefont {M\"{u}hlschlegel}\ \emph {et~al.}(2005)\citenamefont
  {M\"{u}hlschlegel}, \citenamefont {Eisler}, \citenamefont {Martin},
  \citenamefont {Hecht}, \citenamefont {Pohl},\ and\ \citenamefont
  {Muhlschlegel}}]{Muhlschlegel2005}%
  \BibitemOpen
  \bibfield  {author} {\bibinfo {author} {\bibfnamefont {P.}~\bibnamefont
  {M\"{u}hlschlegel}}, \bibinfo {author} {\bibfnamefont {H.-J. H.-J.~J.}\
  \bibnamefont {Eisler}}, \bibinfo {author} {\bibfnamefont {O.~J.~F.}\
  \bibnamefont {Martin}}, \bibinfo {author} {\bibfnamefont {B.}~\bibnamefont
  {Hecht}}, \bibinfo {author} {\bibfnamefont {D.~W.}\ \bibnamefont {Pohl}}, \
  and\ \bibinfo {author} {\bibfnamefont {P.}~\bibnamefont {Muhlschlegel}},\
  }\href {\doibase 10.1126/science.1111886} {\bibfield  {journal} {\bibinfo
  {journal} {Science}\ }\textbf {\bibinfo {volume} {308}},\ \bibinfo {pages}
  {1607 } (\bibinfo {year} {2005})}\BibitemShut {NoStop}%
\bibitem [{\citenamefont {Curto}\ \emph {et~al.}(2010)\citenamefont {Curto},
  \citenamefont {Volpe}, \citenamefont {Taminiau}, \citenamefont {Kreuzer},
  \citenamefont {Quidant},\ and\ \citenamefont {van Hulst}}]{Curto2010}%
  \BibitemOpen
  \bibfield  {author} {\bibinfo {author} {\bibfnamefont {A.~G.}\ \bibnamefont
  {Curto}}, \bibinfo {author} {\bibfnamefont {G.}~\bibnamefont {Volpe}},
  \bibinfo {author} {\bibfnamefont {T.~H.}\ \bibnamefont {Taminiau}}, \bibinfo
  {author} {\bibfnamefont {M.~P.}\ \bibnamefont {Kreuzer}}, \bibinfo {author}
  {\bibfnamefont {R.}~\bibnamefont {Quidant}}, \ and\ \bibinfo {author}
  {\bibfnamefont {N.~F.}\ \bibnamefont {van Hulst}},\ }\href {\doibase
  10.1126/science.1191922} {\bibfield  {journal} {\bibinfo  {journal}
  {Science}\ }\textbf {\bibinfo {volume} {329}},\ \bibinfo {pages} {930}
  (\bibinfo {year} {2010})}\BibitemShut {NoStop}%
\bibitem [{\citenamefont {Dorfm{\"{u}}ller}\ \emph {et~al.}(2010)\citenamefont
  {Dorfm{\"{u}}ller}, \citenamefont {Vogelgesang}, \citenamefont {Khunsin},
  \citenamefont {Rockstuhl}, \citenamefont {Etrich},\ and\ \citenamefont
  {Kern}}]{Dorfmuller2010}%
  \BibitemOpen
  \bibfield  {author} {\bibinfo {author} {\bibfnamefont {J.}~\bibnamefont
  {Dorfm{\"{u}}ller}}, \bibinfo {author} {\bibfnamefont {R.}~\bibnamefont
  {Vogelgesang}}, \bibinfo {author} {\bibfnamefont {W.}~\bibnamefont
  {Khunsin}}, \bibinfo {author} {\bibfnamefont {C.}~\bibnamefont {Rockstuhl}},
  \bibinfo {author} {\bibfnamefont {C.}~\bibnamefont {Etrich}}, \ and\ \bibinfo
  {author} {\bibfnamefont {K.}~\bibnamefont {Kern}},\ }\href {\doibase
  10.1021/nl101921y} {\bibfield  {journal} {\bibinfo  {journal} {Nano letters}\
  }\textbf {\bibinfo {volume} {10}},\ \bibinfo {pages} {3596} (\bibinfo {year}
  {2010})}\BibitemShut {NoStop}%
\bibitem [{\citenamefont {Sauvan}\ \emph {et~al.}(2013)\citenamefont {Sauvan},
  \citenamefont {Hugonin}, \citenamefont {Maksymov},\ and\ \citenamefont
  {Lalanne}}]{Sauvan2013}%
  \BibitemOpen
  \bibfield  {author} {\bibinfo {author} {\bibfnamefont {C.}~\bibnamefont
  {Sauvan}}, \bibinfo {author} {\bibfnamefont {J.~P.}\ \bibnamefont {Hugonin}},
  \bibinfo {author} {\bibfnamefont {I.~S.}\ \bibnamefont {Maksymov}}, \ and\
  \bibinfo {author} {\bibfnamefont {P.}~\bibnamefont {Lalanne}},\ }\href
  {\doibase 10.1103/PhysRevLett.110.237401} {\bibfield  {journal} {\bibinfo
  {journal} {Physical Review Letters}\ }\textbf {\bibinfo {volume} {110}},\
  \bibinfo {pages} {237401} (\bibinfo {year} {2013})}\BibitemShut {NoStop}%
\bibitem [{\citenamefont {Kristensen}\ and\ \citenamefont
  {Hughes}(2014)}]{Kristensen2014}%
  \BibitemOpen
  \bibfield  {author} {\bibinfo {author} {\bibfnamefont {P.~T.~s.}\
  \bibnamefont {Kristensen}}\ and\ \bibinfo {author} {\bibfnamefont
  {S.}~\bibnamefont {Hughes}},\ }\href {\doibase 10.1021/ph400114e} {\bibfield
  {journal} {\bibinfo  {journal} {ACS Photonics}\ }\textbf {\bibinfo {volume}
  {1}},\ \bibinfo {pages} {2} (\bibinfo {year} {2014})}\BibitemShut {NoStop}%
\bibitem [{\citenamefont {Greffet}\ \emph {et~al.}(2010)\citenamefont
  {Greffet}, \citenamefont {Laroche},\ and\ \citenamefont
  {Marquier}}]{Greffet2010}%
  \BibitemOpen
  \bibfield  {author} {\bibinfo {author} {\bibfnamefont {J.-J.}\ \bibnamefont
  {Greffet}}, \bibinfo {author} {\bibfnamefont {M.}~\bibnamefont {Laroche}}, \
  and\ \bibinfo {author} {\bibfnamefont {F.}~\bibnamefont {Marquier}},\ }\href
  {\doibase 10.1103/PhysRevLett.105.117701} {\bibfield  {journal} {\bibinfo
  {journal} {Physical Review Letters}\ }\textbf {\bibinfo {volume} {105}},\
  \bibinfo {pages} {1} (\bibinfo {year} {2010})}\BibitemShut {NoStop}%
\bibitem [{\citenamefont {Krasnok}\ \emph {et~al.}(2015)\citenamefont
  {Krasnok}, \citenamefont {Slobozhanyuk}, \citenamefont {Simovski},
  \citenamefont {Tretyakov}, \citenamefont {Poddubny}, \citenamefont
  {Miroshnichenko}, \citenamefont {Kivshar},\ and\ \citenamefont
  {Belov}}]{Krasnok2015}%
  \BibitemOpen
  \bibfield  {author} {\bibinfo {author} {\bibfnamefont {A.~E.}\ \bibnamefont
  {Krasnok}}, \bibinfo {author} {\bibfnamefont {A.~P.}\ \bibnamefont
  {Slobozhanyuk}}, \bibinfo {author} {\bibfnamefont {C.~R.}\ \bibnamefont
  {Simovski}}, \bibinfo {author} {\bibfnamefont {S.~A.}\ \bibnamefont
  {Tretyakov}}, \bibinfo {author} {\bibfnamefont {A.~N.}\ \bibnamefont
  {Poddubny}}, \bibinfo {author} {\bibfnamefont {A.~E.}\ \bibnamefont
  {Miroshnichenko}}, \bibinfo {author} {\bibfnamefont {Y.~S.}\ \bibnamefont
  {Kivshar}}, \ and\ \bibinfo {author} {\bibfnamefont {P.~A.}\ \bibnamefont
  {Belov}},\ }\href {\doibase 10.1038/srep12956} {\bibfield  {journal}
  {\bibinfo  {journal} {Scientific reports}\ }\textbf {\bibinfo {volume} {5}},\
  \bibinfo {pages} {12956} (\bibinfo {year} {2015})}\BibitemShut {NoStop}%
\bibitem [{\citenamefont {Hecht}\ and\ \citenamefont
  {Novotny}(2012)}]{Hecht2012}%
  \BibitemOpen
  \bibfield  {author} {\bibinfo {author} {\bibfnamefont {B.}~\bibnamefont
  {Hecht}}\ and\ \bibinfo {author} {\bibfnamefont {L.}~\bibnamefont
  {Novotny}},\ }\href@noop {} {\emph {\bibinfo {title} {Principles of
  Nano-Optics}}},\ \bibinfo {edition} {2nd}\ ed.\ (\bibinfo  {publisher}
  {Cambridge},\ \bibinfo {year} {2012})\BibitemShut {NoStop}%
\bibitem [{\citenamefont {Schwinger}\ \emph {et~al.}(1998)\citenamefont
  {Schwinger}, \citenamefont {Deraad},\ and\ \citenamefont
  {Milton}}]{Schwinger1998}%
  \BibitemOpen
  \bibfield  {author} {\bibinfo {author} {\bibfnamefont {J.}~\bibnamefont
  {Schwinger}}, \bibinfo {author} {\bibfnamefont {L.~L.~J.}\ \bibnamefont
  {Deraad}}, \ and\ \bibinfo {author} {\bibfnamefont {K.~A.}\ \bibnamefont
  {Milton}},\ }\href
  {https://books.google.de/books/about/Classical\_Electrodynamics.html?id=SVz\_lRkP09kC\&pgis=1}
  {\emph {\bibinfo {title} {{Classical Electrodynamics - Chpt. 12}}}}\
  (\bibinfo  {publisher} {Westview Press},\ \bibinfo {year} {1998})\ p.\
  \bibinfo {pages} {592}\BibitemShut {NoStop}%
\bibitem [{\citenamefont {Snyder}\ and\ \citenamefont
  {Love}(1983)}]{Snyder1983}%
  \BibitemOpen
  \bibfield  {author} {\bibinfo {author} {\bibfnamefont {A.~A.~W.}\
  \bibnamefont {Snyder}}\ and\ \bibinfo {author} {\bibfnamefont {J.~D.}\
  \bibnamefont {Love}},\ }\href {\doibase 10.1007/978-1-4613-2813-1} {\emph
  {\bibinfo {title} {{Optical waveguide theory}}}},\ Science paperbacks\
  (\bibinfo  {publisher} {Springer US},\ \bibinfo {address} {London ; New
  York},\ \bibinfo {year} {1983})\ p.\ \bibinfo {pages} {734}\BibitemShut
  {NoStop}%
\bibitem [{\citenamefont {Then}\ \emph {et~al.}(2014)\citenamefont {Then},
  \citenamefont {Razinskas}, \citenamefont {Feichtner}, \citenamefont {Haas},
  \citenamefont {Wild}, \citenamefont {Bellini}, \citenamefont {Osellame},
  \citenamefont {Cerullo},\ and\ \citenamefont {Hecht}}]{Then2014}%
  \BibitemOpen
  \bibfield  {author} {\bibinfo {author} {\bibfnamefont {P.}~\bibnamefont
  {Then}}, \bibinfo {author} {\bibfnamefont {G.}~\bibnamefont {Razinskas}},
  \bibinfo {author} {\bibfnamefont {T.}~\bibnamefont {Feichtner}}, \bibinfo
  {author} {\bibfnamefont {P.}~\bibnamefont {Haas}}, \bibinfo {author}
  {\bibfnamefont {A.}~\bibnamefont {Wild}}, \bibinfo {author} {\bibfnamefont
  {N.}~\bibnamefont {Bellini}}, \bibinfo {author} {\bibfnamefont
  {R.}~\bibnamefont {Osellame}}, \bibinfo {author} {\bibfnamefont
  {G.}~\bibnamefont {Cerullo}}, \ and\ \bibinfo {author} {\bibfnamefont
  {B.}~\bibnamefont {Hecht}},\ }\href {\doibase 10.1103/PhysRevA.89.053801}
  {\bibfield  {journal} {\bibinfo  {journal} {Physical Review A}\ }\textbf
  {\bibinfo {volume} {89}},\ \bibinfo {pages} {053801} (\bibinfo {year}
  {2014})}\BibitemShut {NoStop}%
\bibitem [{\citenamefont {Feichtner}\ \emph {et~al.}(2015)\citenamefont
  {Feichtner}, \citenamefont {Selig},\ and\ \citenamefont
  {Hecht}}]{Feichtner2015}%
  \BibitemOpen
  \bibfield  {author} {\bibinfo {author} {\bibfnamefont {T.}~\bibnamefont
  {Feichtner}}, \bibinfo {author} {\bibfnamefont {O.}~\bibnamefont {Selig}}, \
  and\ \bibinfo {author} {\bibfnamefont {B.}~\bibnamefont {Hecht}},\
  }\href@noop {} {\bibfield  {journal} {\bibinfo  {journal} {submitted}\ ,\
  \bibinfo {pages} {16}} (\bibinfo {year} {2015})}\BibitemShut {NoStop}%
\bibitem [{\citenamefont {Garc\'{\i}a-Etxarri}\ \emph
  {et~al.}(2012)\citenamefont {Garc\'{\i}a-Etxarri}, \citenamefont {Apell},
  \citenamefont {K\"{a}ll},\ and\ \citenamefont
  {Aizpurua}}]{Garcia-Etxarri2012}%
  \BibitemOpen
  \bibfield  {author} {\bibinfo {author} {\bibfnamefont {A.}~\bibnamefont
  {Garc\'{\i}a-Etxarri}}, \bibinfo {author} {\bibfnamefont {P.}~\bibnamefont
  {Apell}}, \bibinfo {author} {\bibfnamefont {M.}~\bibnamefont {K\"{a}ll}}, \
  and\ \bibinfo {author} {\bibfnamefont {J.}~\bibnamefont {Aizpurua}},\ }\href
  {\doibase 10.1364/OE.20.025201} {\bibfield  {journal} {\bibinfo  {journal}
  {Optics express}\ }\textbf {\bibinfo {volume} {20}},\ \bibinfo {pages}
  {25201} (\bibinfo {year} {2012})}\BibitemShut {NoStop}%
\bibitem [{\citenamefont {Regmi}\ \emph {et~al.}(2015)\citenamefont {Regmi},
  \citenamefont {{Al Balushi}}, \citenamefont {Rigneault}, \citenamefont
  {Gordon},\ and\ \citenamefont {Wenger}}]{Regmi2015}%
  \BibitemOpen
  \bibfield  {author} {\bibinfo {author} {\bibfnamefont {R.}~\bibnamefont
  {Regmi}}, \bibinfo {author} {\bibfnamefont {A.~A.}\ \bibnamefont {{Al
  Balushi}}}, \bibinfo {author} {\bibfnamefont {H.}~\bibnamefont {Rigneault}},
  \bibinfo {author} {\bibfnamefont {R.}~\bibnamefont {Gordon}}, \ and\ \bibinfo
  {author} {\bibfnamefont {J.}~\bibnamefont {Wenger}},\ }\href {\doibase
  10.1038/srep15852} {\bibfield  {journal} {\bibinfo  {journal} {Scientific
  reports}\ }\textbf {\bibinfo {volume} {5}},\ \bibinfo {pages} {15852}
  (\bibinfo {year} {2015})}\BibitemShut {NoStop}%
\bibitem [{\citenamefont {Schl\"{u}cker}(2014)}]{Schlucker2014}%
  \BibitemOpen
  \bibfield  {author} {\bibinfo {author} {\bibfnamefont {S.}~\bibnamefont
  {Schl\"{u}cker}},\ }\href {\doibase 10.1002/anie.201205748} {\bibfield
  {journal} {\bibinfo  {journal} {Angewandte Chemie (International ed. in
  English)}\ }\textbf {\bibinfo {volume} {53}},\ \bibinfo {pages} {4756}
  (\bibinfo {year} {2014})}\BibitemShut {NoStop}%
\bibitem [{\citenamefont {Drexhage}(1974)}]{Drexhage1974}%
  \BibitemOpen
  \bibfield  {author} {\bibinfo {author} {\bibfnamefont {K.~H.}\ \bibnamefont
  {Drexhage}},\ }in\ \href {\doibase 10.1016/S0079-6638(08)70266-X} {\emph
  {\bibinfo {booktitle} {Progress in Optics XII}}},\ \bibinfo {editor} {edited
  by\ \bibinfo {editor} {\bibfnamefont {E.}~\bibnamefont {Wolf}}}\ (\bibinfo
  {publisher} {Elsevier},\ \bibinfo {year} {1974})\ pp.\ \bibinfo {pages} {163
  -- 232}\BibitemShut {NoStop}%
\bibitem [{\citenamefont {Olmon}\ and\ \citenamefont
  {Raschke}(2012)}]{Olmon2012}%
  \BibitemOpen
  \bibfield  {author} {\bibinfo {author} {\bibfnamefont {R.}~\bibnamefont
  {Olmon}}\ and\ \bibinfo {author} {\bibfnamefont {M.}~\bibnamefont
  {Raschke}},\ }\href {\doibase 10.1088/0957-4484/23/44/444001} {\bibfield
  {journal} {\bibinfo  {journal} {Nanotechnology}\ }\textbf {\bibinfo {volume}
  {23}},\ \bibinfo {pages} {444001} (\bibinfo {year} {2012})}\BibitemShut
  {NoStop}%
\bibitem [{Note1()}]{Note1}%
  \BibitemOpen
  \bibinfo {note} {The mode and the spectra were obtained numerically using
  Lumerical FDTD Solutions, using focussed broadband ($\lambda = 550 - 750$ nm)
  Gaussian illumination ($NA=1$, $3.9$ fs pulse length). The gold material was
  implemented according to\cite {Etchegoin2006}}\BibitemShut {NoStop}%
\bibitem [{\citenamefont {Lee}(1984)}]{Lee1984}%
  \BibitemOpen
  \bibfield  {author} {\bibinfo {author} {\bibfnamefont {K.~F.}\ \bibnamefont
  {Lee}},\ }\href@noop {} {\emph {\bibinfo {title} {Principles of antenna
  theory}}}\ (\bibinfo  {publisher} {John Wiley \& Sons Ltd},\ \bibinfo {year}
  {1984})\BibitemShut {NoStop}%
\bibitem [{\citenamefont {Tretyakov}(2014)}]{Tretyakov2014}%
  \BibitemOpen
  \bibfield  {author} {\bibinfo {author} {\bibfnamefont {S.}~\bibnamefont
  {Tretyakov}},\ }\href {\doibase 10.1007/s11468-014-9699-y} {\bibfield
  {journal} {\bibinfo  {journal} {Plasmonics}\ }\textbf {\bibinfo {volume}
  {9}},\ \bibinfo {pages} {935} (\bibinfo {year} {2014})}\BibitemShut {NoStop}%
\bibitem [{\citenamefont {Etchegoin}\ \emph {et~al.}(2006)\citenamefont
  {Etchegoin}, \citenamefont {{Le Ru}},\ and\ \citenamefont
  {Meyer}}]{Etchegoin2006}%
  \BibitemOpen
  \bibfield  {author} {\bibinfo {author} {\bibfnamefont {P.~G.}\ \bibnamefont
  {Etchegoin}}, \bibinfo {author} {\bibfnamefont {E.~C.}\ \bibnamefont {{Le
  Ru}}}, \ and\ \bibinfo {author} {\bibfnamefont {M.}~\bibnamefont {Meyer}},\
  }\href {\doibase doi:10.1063/1.2360270} {\bibfield  {journal} {\bibinfo
  {journal} {The Journal of Chemical Physics}\ }\textbf {\bibinfo {volume}
  {125}},\ \bibinfo {pages} {164705} (\bibinfo {year} {2006})}\BibitemShut
  {NoStop}%
\bibitem [{\citenamefont {Ruppin}(1982)}]{Ruppin1982}%
  \BibitemOpen
  \bibfield  {author} {\bibinfo {author} {\bibfnamefont {R.}~\bibnamefont
  {Ruppin}},\ }\href {\doibase doi:10.1063/1.443196} {\bibfield  {journal}
  {\bibinfo  {journal} {The Journal of Chemical Physics}\ }\textbf {\bibinfo
  {volume} {76}},\ \bibinfo {pages} {1681} (\bibinfo {year}
  {1982})}\BibitemShut {NoStop}%
\bibitem [{\citenamefont {Kerker}\ \emph {et~al.}(1980)\citenamefont {Kerker},
  \citenamefont {Wang},\ and\ \citenamefont {Chew}}]{Kerker1980}%
  \BibitemOpen
  \bibfield  {author} {\bibinfo {author} {\bibfnamefont {M.}~\bibnamefont
  {Kerker}}, \bibinfo {author} {\bibfnamefont {D.-S.}\ \bibnamefont {Wang}}, \
  and\ \bibinfo {author} {\bibfnamefont {H.}~\bibnamefont {Chew}},\ }\href
  {\doibase 10.1364/AO.19.004159} {\bibfield  {journal} {\bibinfo  {journal}
  {Applied Optics}\ }\textbf {\bibinfo {volume} {19}},\ \bibinfo {pages} {4159}
  (\bibinfo {year} {1980})}\BibitemShut {NoStop}%
\bibitem [{\citenamefont {Mie}(1908)}]{Mie1908}%
  \BibitemOpen
  \bibfield  {author} {\bibinfo {author} {\bibfnamefont {G.}~\bibnamefont
  {Mie}},\ }\href {\doibase 10.1002/andp.19083300302} {\bibfield  {journal}
  {\bibinfo  {journal} {Annalen der Physik}\ }\textbf {\bibinfo {volume}
  {330}},\ \bibinfo {pages} {377–445} (\bibinfo {year} {1908})}\BibitemShut
  {NoStop}%
\bibitem [{\citenamefont {Carminati}\ \emph {et~al.}(2006)\citenamefont
  {Carminati}, \citenamefont {Greffet}, \citenamefont {Henkel},\ and\
  \citenamefont {Vigoureux}}]{Carminati2006}%
  \BibitemOpen
  \bibfield  {author} {\bibinfo {author} {\bibfnamefont {R.}~\bibnamefont
  {Carminati}}, \bibinfo {author} {\bibfnamefont {J.~J.}\ \bibnamefont
  {Greffet}}, \bibinfo {author} {\bibfnamefont {C.}~\bibnamefont {Henkel}}, \
  and\ \bibinfo {author} {\bibfnamefont {J.~M.}\ \bibnamefont {Vigoureux}},\
  }\href {\doibase 10.1016/j.optcom.2005.12.009} {\bibfield  {journal}
  {\bibinfo  {journal} {Optics Communications}\ }\textbf {\bibinfo {volume}
  {261}},\ \bibinfo {pages} {368} (\bibinfo {year} {2006})}\BibitemShut
  {NoStop}%
\bibitem [{\citenamefont {Bohren}\ and\ \citenamefont
  {Huffman}(1983)}]{Bohren1983}%
  \BibitemOpen
  \bibfield  {author} {\bibinfo {author} {\bibfnamefont {C.~F.}\ \bibnamefont
  {Bohren}}\ and\ \bibinfo {author} {\bibfnamefont {D.~R.}\ \bibnamefont
  {Huffman}},\ }\href@noop {} {\emph {\bibinfo {title} {Absorption and
  scattering of light by small particles}}}\ (\bibinfo  {publisher} {Wiley},\
  \bibinfo {year} {1983})\ p.\ \bibinfo {pages} {530}\BibitemShut {NoStop}%
\bibitem [{Note2()}]{Note2}%
  \BibitemOpen
  \bibinfo {note} {Ruppin\cite {Ruppin1982} missed the prefactor $D_\nu $,
  which is correctly included in the work of Kerker\cite
  {Kerker1980}}\BibitemShut {NoStop}%
\bibitem [{\citenamefont {Doyle}(1989)}]{Doyle1989}%
  \BibitemOpen
  \bibfield  {author} {\bibinfo {author} {\bibfnamefont {W.~T.}\ \bibnamefont
  {Doyle}},\ }\href {\doibase 10.1103/PhysRevB.39.9852} {\bibfield  {journal}
  {\bibinfo  {journal} {Physical Review B}\ }\textbf {\bibinfo {volume} {39}},\
  \bibinfo {pages} {9852} (\bibinfo {year} {1989})}\BibitemShut {NoStop}%
\bibitem [{\citenamefont {Feichtner}\ \emph {et~al.}(2012)\citenamefont
  {Feichtner}, \citenamefont {Selig}, \citenamefont {Kiunke},\ and\
  \citenamefont {Hecht}}]{Feichtner2012}%
  \BibitemOpen
  \bibfield  {author} {\bibinfo {author} {\bibfnamefont {T.}~\bibnamefont
  {Feichtner}}, \bibinfo {author} {\bibfnamefont {O.}~\bibnamefont {Selig}},
  \bibinfo {author} {\bibfnamefont {M.}~\bibnamefont {Kiunke}}, \ and\ \bibinfo
  {author} {\bibfnamefont {B.}~\bibnamefont {Hecht}},\ }\href {\doibase
  10.1103/PhysRevLett.109.127701} {\bibfield  {journal} {\bibinfo  {journal}
  {Physical Review Letters}\ }\textbf {\bibinfo {volume} {109}},\ \bibinfo
  {pages} {127701} (\bibinfo {year} {2012})}\BibitemShut {NoStop}%
\bibitem [{\citenamefont {Enderlein}(2002)}]{Enderlein2002}%
  \BibitemOpen
  \bibfield  {author} {\bibinfo {author} {\bibfnamefont {J.}~\bibnamefont
  {Enderlein}},\ }\href {\doibase 10.1063/1.1434314} {\bibfield  {journal}
  {\bibinfo  {journal} {Applied Physics Letters}\ }\textbf {\bibinfo {volume}
  {80}},\ \bibinfo {pages} {315} (\bibinfo {year} {2002})}\BibitemShut
  {NoStop}%
\bibitem [{\citenamefont {Novotny}(2007)}]{Novotny2007}%
  \BibitemOpen
  \bibfield  {author} {\bibinfo {author} {\bibfnamefont {L.}~\bibnamefont
  {Novotny}},\ }\href {\doibase 10.1103/PhysRevLett.98.266802} {\bibfield
  {journal} {\bibinfo  {journal} {Physical Review Letters}\ }\textbf {\bibinfo
  {volume} {98}},\ \bibinfo {pages} {266802} (\bibinfo {year}
  {2007})}\BibitemShut {NoStop}%
\end{thebibliography}%

\section{Supplementary material}

\section{Appendix I: Dipole in front of sphere}
To validate eq.~\eqref{eq:dipTotalPower_reci} we consider the problem of a dipole in front of a sphere with radius $R$. This case has been treated analytically by Kerker \cite{Kerker1980} and Ruppin \cite{Ruppin1982} based on Mie theory \cite{Mie1908}. Later a generalized expression for the emission power enhancement for a dipole close to a small spheres was derived based on the sphere's polarizability \cite{Carminati2006}, which we will reproduce here.

We start by expressing the mode current in terms of the spheres mode fields:
\begin{equation}
\mathbf{j}_\nu = \sigma \mathbf{E}_\nu = i \omega \varepsilon_0 (\varepsilon (\omega) - 1) \mathbf{E}_\nu \, .
\label{eq:mode_current_to_fields}
\end{equation}
Inserting eq.~\eqref{eq:mode_current_to_fields} into eq.~\eqref{eq:dipTotalPower_reci} then leads to:
\begin{multline}
\frac{P}{P_0} = 1 + \frac{6 \pi \varepsilon_0^2}{k^3} \cdot \\ \cdot \text{Im}\, \left \{ (\varepsilon (\omega)-1) \int_{V_\nu} \left | \mathbf{E}_p (\mathbf{r}') \cdot \mathbf{E}_\nu (\mathbf{r}') \right | \, \text{d}^3 r' \right \}.
\label{eq:power_fields}
\end{multline}
The dipole fields inside the sphere volume can be expanded into Mie modes:
\begin{equation}
\mathbf{E}_\text{p}(\mathbf{r},\omega) = \sum_\nu D_\nu \left[ p_\nu \mathbf{M}^{(1)}_\nu (k\mathbf{r}) + q_\nu \mathbf{N}^{(1)}_\nu(k\mathbf{r}) \right]
\label{eq:spherical_scattered_field}
\end{equation}
$\mathbf{M}^{(1)}_\nu$ and $\mathbf{N}^{(1)}_\nu$ are the spherical vector wave functions, as defined in \cite{Bohren1983} for a given set of control variables $\nu = n, m, \sigma$ with $n \in \mathbb{N}, m \leq n \in \mathbb{N}$ and $\sigma =$ odd or even, $D_\nu = \xi [(2n+1)(n-m)!]/[4n(n+1)(n+m)!]$ with $\xi = 1$ if $m=0$ or $\xi = 2$ if $m>0$, and
\begin{align}
p_\nu &= \frac{ik^3}{\varepsilon_0 \pi}\mathbf{M}^{(3)}_\nu(k\mathbf{r}_0) \cdot \mathbf{p} \\
q_\nu &= \frac{ik^3}{\varepsilon_0 \pi}\mathbf{N}^{(3)}_\nu(k\mathbf{r}_0) \cdot \mathbf{p}
\label{eq:DipoleFieldsFactors}
\end{align}
being prefactors originating from the Greens tensor\cite{Kerker1980}.

The mode fields of the sphere read as \footnote{Ruppin\cite{Ruppin1982} missed the prefactor $D_\nu$, which is correctly included in the work of Kerker\cite{Kerker1980}}:
\begin{equation}
\mathbf{E}_\text{sph}(\mathbf{r}) = \sum_\nu D_\nu \left[ f_\nu \mathbf{M}^{(1)}_\nu (k_1\mathbf{r}) + g_\nu \mathbf{N}^{(1)}_\nu(k_1\mathbf{r}) \right]\;  ,
\label{eq:spherical_sphere_field}
\end{equation} 
with $k_1 = k\cdot\sqrt{\varepsilon(\omega)}$, and the factors
\begin{equation}
f_\nu = \alpha_n p_\nu \quad ; \quad 
g_\nu = \beta_n q_\nu \; ,
\label{eq:SphereFieldsFactors}
\end{equation}
where $\alpha_n$ and $\beta_n$ are complex valued Mie-like coefficients which are derived from the boundary conditions for electric fields at the sphere surface \cite{Ruppin1982}. 

For a sphere with small radius and small dipole distances $R \ll r_p \ll \lambda$ we can restrict our calculation to the emission power enhancement due to the fundamental dipolar sphere mode $\mathbf{N}_{1,0,\text{odd}}^{(1)} = \mathbf{N}_p$ leading to:
\begin{multline}
\int_{V_\text{sph}} \mathbf{E}_p \cdot \mathbf{E}_\nu \, \text{d}V =\\
= \int_{V_\text{sph}} \left ( D_1 q_{1,0,\text{odd}} \mathbf{N}_p(k\mathbf{r})\right) \cdot \\ \cdot \left ( D_1 \beta q_{1,0,\text{odd}} \mathbf{N}_p(k_1\mathbf{r})\right) \, \text{d}V\, .
\label{eq:ReciSphere}
\end{multline}
Here we made use of the fact that the spherical vector wave functions are orthogonal
\begin{equation}
\int_{V_\text{sph}} A_\nu \cdot B_\mu^* \;\text{d}V = 0
\label{eq:SVWF_Orthonormality}
\end{equation}
for $A,B \in {\mathbf{N}^{(1)},\mathbf{M}^{(1)}}$ with $A \neq B$ and arbitrary $\nu, \mu$.

In the limit of small spheres the terms of the integral \eqref{eq:ReciSphere} can be developed into series of $kR$ and $kr_p$ respectively, which leads to the following intermediate results:
\begin{align}
q_1 &= \frac{ik^3}{\pi \varepsilon_0} \frac{2}{kr_p} \cdot h_1(kr_p) \\
\text{with }h_1(kr_p) &= e^{ikr_p}\left( \frac{1}{kr_p} - \frac{i}{(kr_p)^2}\right) \\
\beta &= \frac{3}{\varepsilon(\omega) + 2} + \mathcal{O}(k^2R^2)
\label{eq:inter}
\end{align}
\begin{equation}
\int_{V_\text{sph}} \mathbf{N}_p(k\mathbf{r'}) \cdot \mathbf{N}_p(k_1\mathbf{r'}) \, d^3r' = \frac{16}{27}\pi R^3 + \mathcal{O}(R^5)
\label{eq:interInt}
\end{equation}
with $h_1$ being the Hankel function of the first kind. Putting the integral together with $D_1=3/8$ and inserting it into eq.~\eqref{eq:power_fields} leads to the final result:
\begin{multline}
\frac{P}{P_0} = 1 + \frac{3k^3}{2\pi} \cdot \\
\cdot \text{Im} \left\{ \alpha_0(\omega) e^{2ikr_p} \left[ \frac{1}{(kr_p)^4} - \frac{2i}{(kr_p)^5} - \frac{1}{(kr_p)^6}\right]\right\}\; ,
\label{eq:Result_Sphere}
\end{multline}
which is identical to the result in \cite{Carminati2006}, yet with $\alpha_0 = 4\pi R^3 (\varepsilon(\omega) - 1)/(\varepsilon(\omega)+2)$ being the quasi-static polarizability of a small sphere. Taking more terms of $\beta$ into account the same result can be derived for the effective polarizability including also radiative losses \cite{Doyle1989}.

\section{Appendix II: Revisiting the split-ring-antenna}\label{app:2}
\begin{figure}[tp]%
\centering
\includegraphics[scale=1]{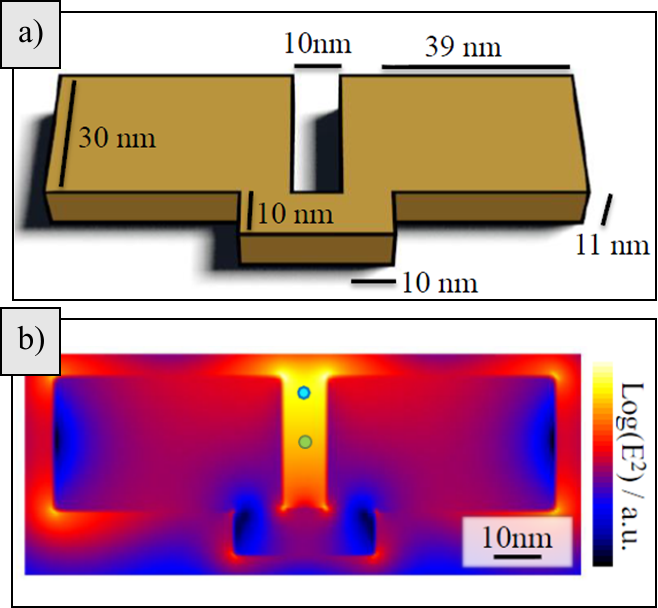}
\caption{\label{fig:SRA}(color online) Evaluation of a split-ring antenna via the mode-matching method. (a) geometry of split-ring antenna as introduced in \cite{Feichtner2012}. (b) Near-field intensity distribution in the center $x$-$y$-plane. The green and blue dot represent the center of the antenna and the position of highest fields along the $y$-axis, respectively (more information in text).}%
\end{figure}

In \cite{Feichtner2012} the split-ring-antenna (sketched in Fig.~\ref{fig:SRA}(a)) was introduced as a result of evolutionary optimization. It has been shown that it outperforms a comparable two-wire dipolar nano antenna, reasoned by the additional current from the shortcut across the antenna gap enabling a split-ring like mode, which adds up constructively with the dipolar antenna currents for charge accumulation at the gap. Using the mode matching formalism it can now be understand, that the short cut adds a current path resembling dipolar fields in the very center of the split ring antenna.

We use this system for a numerical test of the new theory, as the SRA shows only one excitable antenna mode. We chose two positions for the dipole, one in the very center of the gap (green circle), one in the point of highest fields along the $y$-axis in 11 nm distance from the center (blue circle). The ratio of mode near-field at the given positions is evaluated by illuminating the antenna with a normalized Gaussian of NA$=1$ and $\lambda=650$ nm, yielding a power ratio of $P_\text{rel}=E_\text{max}/E_\text{center}=1.441$. To check the validity of eq.~\eqref{eq:dipTotalPower_reci}, it is integrated numerically to obtain the mode-matching power enhancement factor $P_{mm}$:
\begin{equation}
P_\text{mm} = \frac{P}{P_0} \propto \sum_r \mathbf{E}_\nu(r) \cdot \mathbf{E}_\text{dip}(r) \quad .
\label{eq:overlapSum}
\end{equation}
Here $r$ indexes all Yee-Cells within the antenna volume. The antenna has a large enough field enhancement, that direct far-field emission can be neglected. $\mathbf{E}_\nu$ was calculated with the above mentioned Gaussian excitation, the two $\mathbf{E}_\text{dip}$ with a dipole source at a center frequency $\lambda_\text{dip}=650$ nm and a pulse length of $\approx 4$ fs at the respective positions and a subsequent Fourier-transformation to retrieve the quasi-static fields. The ratio $P_\text{mm,max}/P_\text{mm,center}=1.451$ differs from $P_\text{rel}$ only by a factor of 0.007 which is within any error margin due to numerical inaccuracies for the different light source setups.

\section{Appendix III: Field distortion at the air-gold interface}

\begin{figure}
\centering
\includegraphics[width = 0.8\columnwidth]{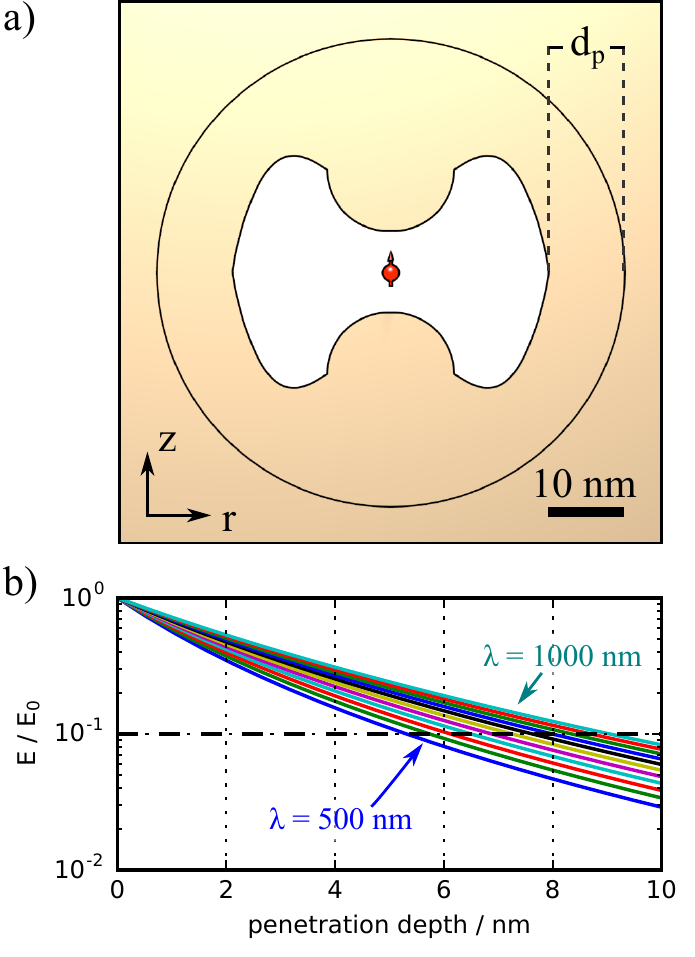}
\caption{(color online) The rotational symmetric geometry for the quasistatic test of field distortion. The center is filled with a single point dipole situated in the very center of the void made from vacuum. The surrounding is made from gold. The circle denotes the region, in which eq.~\eqref{eq:overlap} is evaluated. The minimal distance $d_p$ was set to 10 nm (see text for details).}
\label{fig:dist1}
\end{figure}

To prove the assumption that at optical frequencies the fields of a dipole are not strongly altered by entering a gold surface, the following figure of merit $\mu$ was devised:
\begin{equation}
\mu = \frac{1}{V}\int{\frac{\mathbf{E}_\text{dip,0}(\mathbf{r})\cdot\mathbf{E}_\text{dip,Au}(\mathbf{r})}{\left|\mathbf{E_\text{dip,0}} (\mathbf{r})\right| \left| \mathbf{E}_\text{dip,Au}(\mathbf{r})\right|}}\, \text{d}V
\label{eq:overlap}
\end{equation}
The numerator in the integral is a scalar product of $\mathbf{E}_\text{dip,0}$ the dipole fields in vacuum and $\mathbf{E}_\text{dip,Au}$ the dipole fields in gold at the same point in space $\mathbf{r}$. Together with the normalization denominator the integrand lies in the interval $[-1,1]$, as does $\mu$. The two fields are retrieved from two different quasistatic simulations, where the field in the gold material was retrieved from a simulation filled with gold, except of a void in the very center shaped identically to the cavity of the plasmonic cavity antenna (see Fig.~2 of the main manuscript) as depicted in Fig.~\ref{fig:dist1}(a). The gold materials complex dielectric $\varepsilon(\lambda) = \varepsilon'(\lambda) + i\, \varepsilon''(\lambda)$ was implemented following the Etchegoin model \cite{Etchegoin2006}.

To define the outer border of the integration volume, an estimation for the field penetration was performed assuming the dipole fields with a $r^{-3}$-distance dependence being additionally damped by a factor $e^{-r/\delta(\lambda)}$ after entering the material, with $\delta$ being the penetration depth of surface plasmon fields on a plane metal/air interface \cite{Hecht2012}:
\begin{equation}
\delta_1(\lambda) = \frac{\lambda}{2\pi} \left(\text{Im}\left(\sqrt{\frac{\varepsilon'(\lambda)^2}{\varepsilon'(\lambda)+1}}\right)\right)^{-\frac{1}{2}}
\label{eq:plasmon_penetration_depth}
\end{equation}
The resulting field decay normalized to the surface field strength for a dipole in a spherical cavity with a radius of $\lambda/100$ for wavelengths from 500 to 1000 nm in steps of 50 nm is depicted in Fig.~\ref{fig:dist1}(b). The minimal distance $d_p$ between the void and the outer integration limit of eq.~\eqref{eq:overlap} is set to $d_p = 10$ nm, leading to an overall radius of the integrated volume of 30 nm to ensure that all fields decayed to $< 0.1$ of the surface field strength.

\begin{table}
\centering
\begin{tabular}{|l|c|c|}
\hline
   $\lambda$/nm &       $\mu$ & $\mu_\text{Sphere}$\\
\hline
         500 &  0.242 & -1.000\\
         550 & -0.840 & -0.999\\
         600 & -0.861 & -0.999\\
         650 & -0.878 &	-0.998\\
         700 & -0.887 &	-0.998\\
         750 & -0.892 &	-0.998\\
         800 & -0.896 &	-0.997\\
         850 & -0.898 &	-0.997\\
         900 & -0.900 &	-0.996\\
         950 & -0.902 & -0.996\\
        1000 & -0.903 & -0.996\\
\hline
\end{tabular}
\vspace{0.2cm}
\caption{Field distortion for different dipole emission wavelengths $\lambda$. $\mu$ is the value for the plasmonic cavity antenna center geometry as depicted in Fig.~\ref{fig:dist1}(a), while $\mu_\text{sphere}$ is the same calculation performed for a dipole in a spherical void (details in text).}
\label{tab:mus}
\end{table}

\begin{figure}[ht]
	\centering
		\includegraphics[width = \columnwidth]{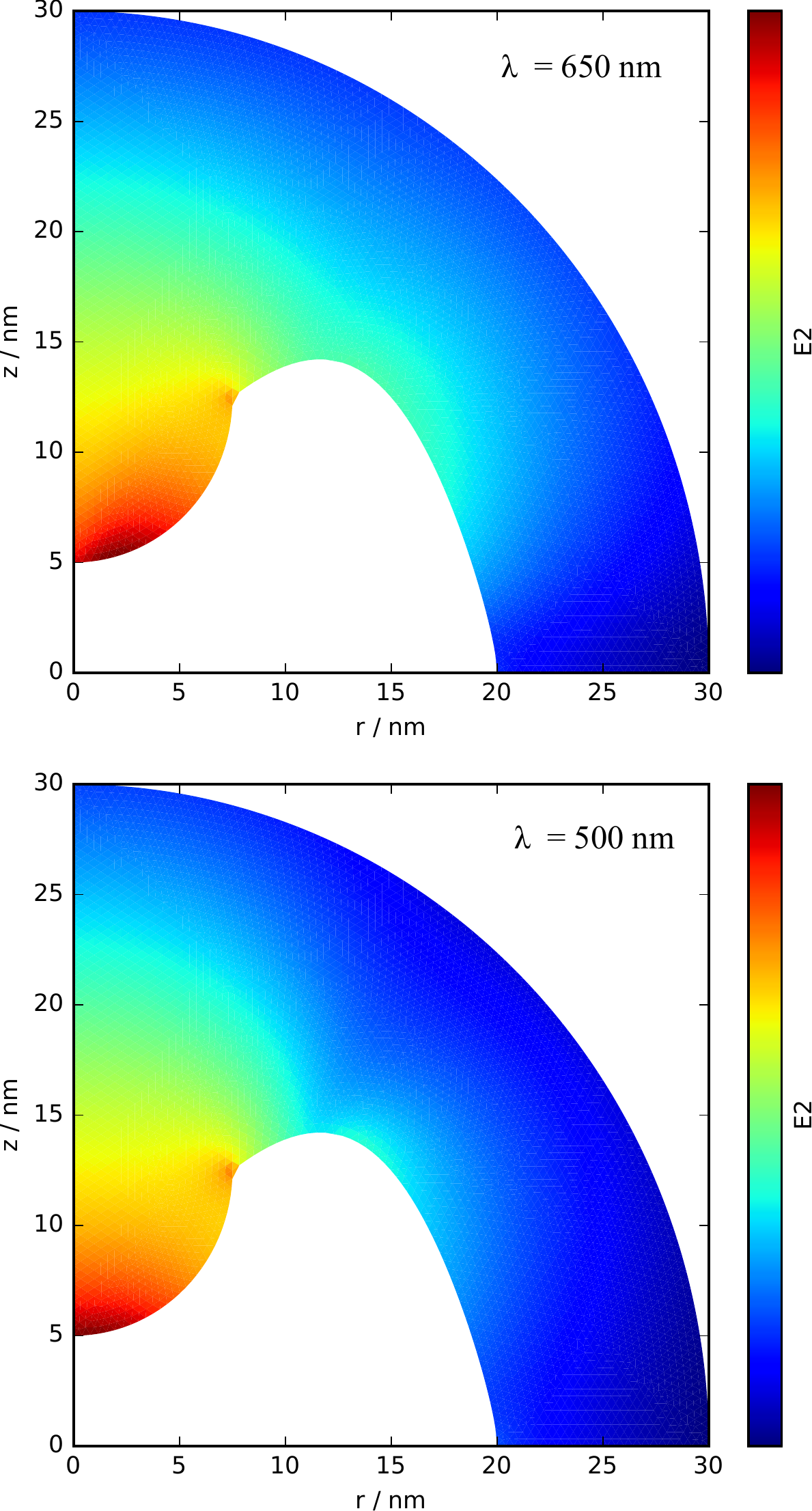}
	\caption{Quasistatic near field intensity for $\lambda = 650$ nm (upper panel) and $\lambda = 500$ nm (lower panel) in the center region of the plasmonic cavity antenna, which was used to evaluate eq.~\eqref{eq:overlap}. The plot shows a quarter of the cross section incorporating the axis of rotation. For $\lambda = 500$ nm an additional field strength minimum has appeared, as a higher order mode is getting excited.}
	\label{fig:Fig_fields}
\end{figure}

The resulting $\mu$ for a set of wavelengths $\lambda$ is given in Table \ref{tab:mus}. Obviously, the values are near to -1 which is expected since the fields acquire a phase shift of $\pi$ on entering a metal. The absolute values are large enough to justify the assumption of nearly unperturbed dipole fields and the mode patterns shown in Fig.~\ref{fig:int_design}(a) can therefore be used as design guidelines. Finally, for $\lambda = 500$ nm the absolute value of $\mu$ drops by a large margin. This is the results of the mode between the two cavity tips switching from a mode with no field node to one with two field nodes as can be seen in Fig.~\ref{fig:Fig_fields}. A non-negligible portion of the field is now in phase with the excitation.

It has to be mentioned, that for the cavity shape being spherical, $\mu_\text{Sphere} < -0.995$ is true for all examined wavelengths, when the sphere radius is set to $r=\lambda/100$ and $d_p = 10$ nm. This originates from the cavity showing the identical symmetry as the dipolar fields. However, the overlap integral and therefore the power transfer is also dependent on the field strength, which can for long wavelengths be optimized by tips and their corresponding lightning rod effect. For short wavelengths near the plasma frequency the metallic behavior gets less and less pronounced and a spherical cavity with the lowest possible radius seems to ensure optimal coupling.

\section{Appendix IV: N-type mode plasmonic cavity antenna}

\begin{figure}[ht]
	\centering
		\includegraphics[width=\columnwidth]{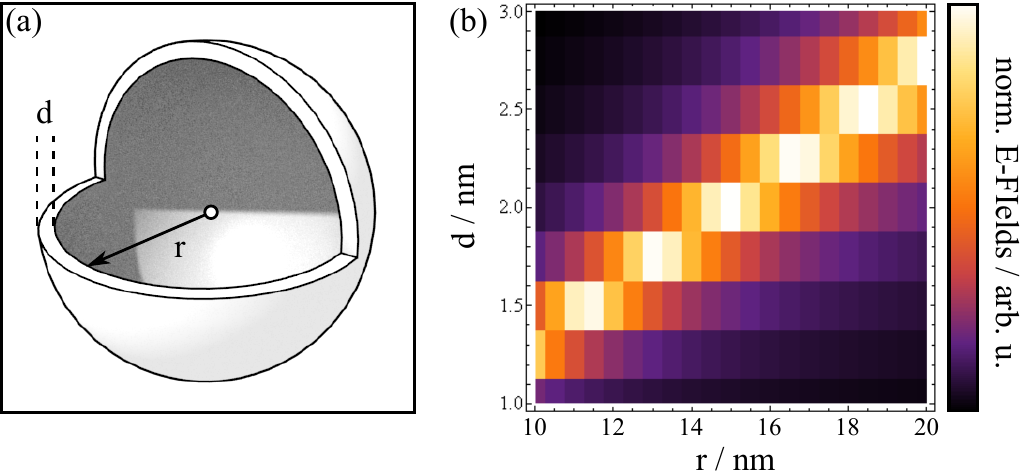}
	\caption{Resonances of sphere shells. (a) Sketch of an air filled metal shell, with cutaway for better visibility of the dimensions $r$, the inner radius and $d$, the shell thickness. (b) Normalized electric fields in the very center (white circle) of a gold nanoshell for a quasi-static excitation at $\lambda = 650$ nm for changing shell radius and thickness.}
	\label{fig:shell_resonance}
\end{figure}

To realize a plasmonic cavity antenna with cylindrical symmetry that exhibits a the n-type mode current pattern while being similar to the reference antenna, the two antenna wires have to be shortcut close to the feedpoint using a metal bridge shaped in a way that resembles the dipole field loops. To asses the optimal thickness of such a geometry we examine the approximative case of a spherical shell. Fig.~\ref{fig:shell_resonance} shows the near-field intensity enhancement in the center of spherical shells for an incoming plane wave with $\lambda = 650$ nm, while the shell radius $r$ and shell thickness $d$ are varied (compare also to \cite{Enderlein2002}).
For small shell radii with dimensions of typical optical antenna gaps, the resonant shell thickness has to be comparably thin, between 1 and 3 nm. This can be understood by an effective wavelength argument: The geometry in Fig.~\ref{fig:int_design}(b) realizes the gap shortcut in a planar gold sheet with 30 nm thickness via two wires with a 'rectangular' cross section. To transform it to a 3D shortcut, the 'height' of the wire has to increase and its width has to shrink to keep the overall shortcut cross section area about constant to keep the identical effective plasmon wavelength and thus conserve the resonance peak position\cite{Novotny2007}.

Simply combining a plasmonic dipolar antenna with a 10 nm gap with a shell with $r = 10$ nm and $d = 1$ nm results in the desired mode in quasi-static FEM simulations (COMSOL).  Yet, it was not possible to validate the geometry by means of full wave FDTD simulations, as down to a mesh size of 0.25 nm the stair-casing effect of the thin spherical shell leads to a severe mode shift and therefore to different NFIE spectra. Finer meshing was not feasible due to large simulation times. We note that this geometry would be of pure academic value, as a 1 nm thick, smooth spherical gold shell filled with air is far from experimental realization.

\end{document}